*Letter to the Editor*

# Saturn's Rings are Fractal

**Jun Li, Martin Ostoja-Starzewski[1]**

The images recently sent by the Cassini spacecraft mission (available on the NASA website http://saturn.jpl.nasa.gov/photos/halloffame/) show the complex and beautiful rings of Saturn. Beginning with (1-3), there have been conjectures that radial cross-sections of Saturn's rings are Cantor sets, but no convincing fractal analysis of actual images ever appeared. Of the 87 Cassini images, in Fig. 1(A) we reproduce slide #42 bearing the title "Mapping Clumps in Saturn's Rings" and in Fig. 1(C) the slide #54 titled "Scattered Sunshine". The first of these is a false-color image of Saturn's main rings made by combining data from multiple star occultations using the Cassini ultraviolet imaging spectrograph. In the second of these, Saturn's icy rings shine in scattered sunlight, from about 15º above the ring plane. The third is the image sent by 'Voyager 2' spacecraft in 1981 from a distance of 8.9 million km (from the NASA website http://solarsystem.nasa.gov/planets/profile.cfm?Object=Saturn&Display=Rings).

Using the box-counting method, we determine the fractal dimension of those rings to be about 1.6~1.7. Each color image in Fig. 1 is first converted into a gray image. Various edge detection methods are then performed and compared to optimally identify ring boundaries: 'Sobel', 'Robert', 'Laplacian of Gaussian', 'Canny' and 'Zero-Cross' edge functions in the Matlab Image Processing Toolbox. Furthermore, the morphology operation functions of 'bridge' and 'skel' are employed to bridge unconnected pixels and remove extra pixels on the boundaries, respectively, in order to eliminate artificial effects from consideration of reality.

We perform several box counting methods to estimate fractal dimensions of the above processed black-white images of Saturn rings:
1. Modified box counting using boxes with shape being self-similar to the global image, which is well suited for the generally rectangular image (4).
2. Power 2 box counting using boxes with sizes as powers of 2, with optimal log-log regression while the partial boarder effects are evident generally.
3. Divider box counting using boxes with sizes being the dividers of the image size. Subsequent box size may be too close for log-log regression, while the border effects can be eliminated.

The local slopes of log(*N*)-log(*R*) are also acquired to determine optimal cut-offs of box sizes. The cut-offs are specified where the local slope varies strongly. The *log*(*N*)-*log*(*R*) plots of these three methods are then shown in Figs 2~4, respectively. We note that, for modified box counting, *R* denotes the ratio of image size to box size, unlike power 2 or divider box counting, where *R* is the box size. *N* is referred to the number of boxes to cover ring edges with a certain size of box. It is well known that for a fractal object $N \propto R^{-D}$ (1). The fractal dimension *D* comes from estimation of the slope of *log*(*N*)-*log*(*R*).

---

[1] Department of Mechanical Science and Engineering, Institute for Condensed Matter Theory, and Beckman Institute, University of Illinois at Urbana-Champaign, Urbana, IL 61801. Contact for correspondence: martinos@illinois.edu

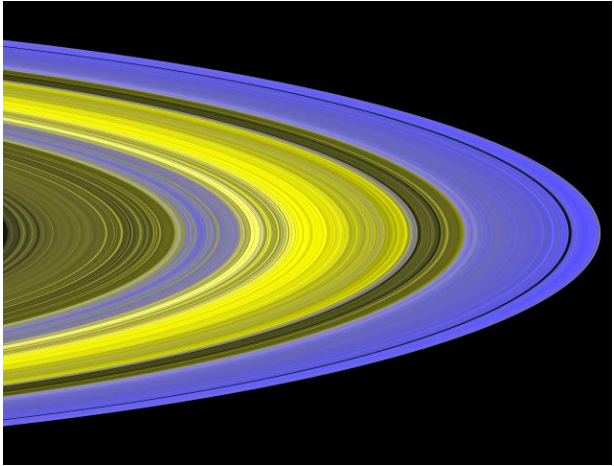

(A)

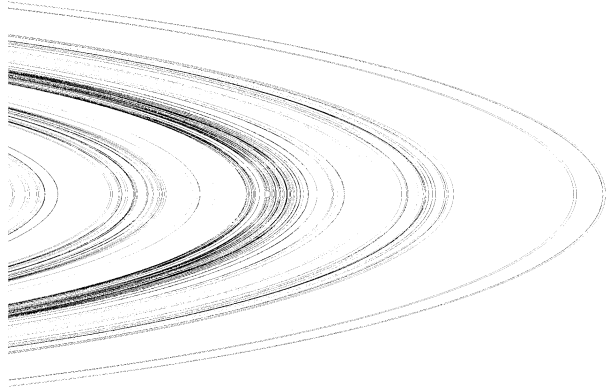

(B)

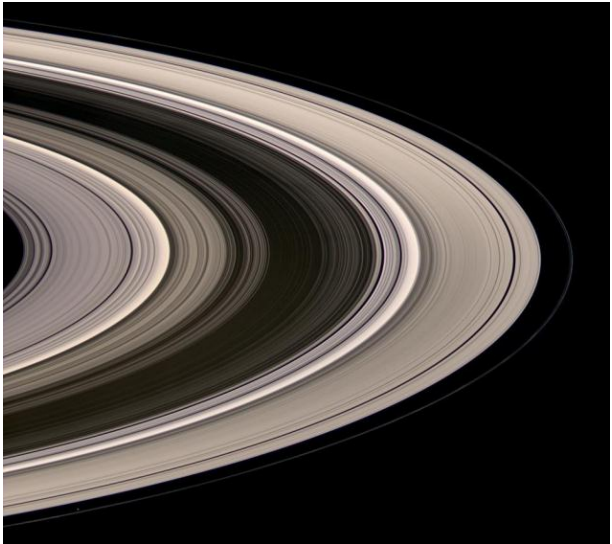

(C)

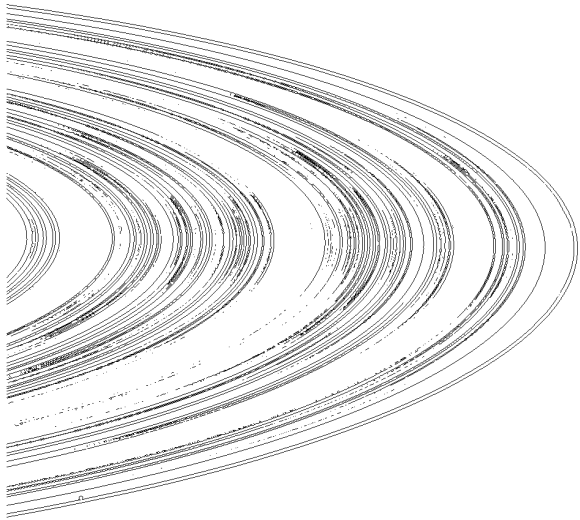

(D)

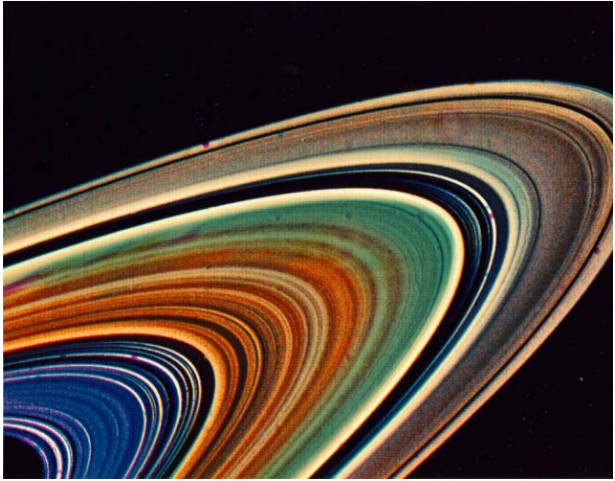 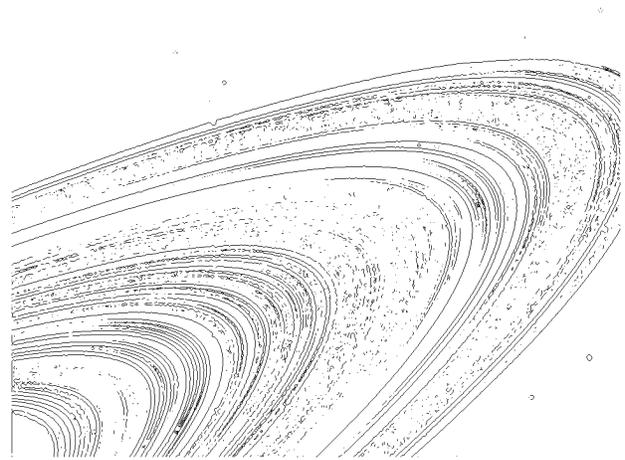

(E) (F)

Fig. 1. (A) The original image 1. (B) Image 1 processed to capture ring edges. (C) The original image 2. (D) Image 2 processed to capture ring edges. (E) The original image 3. (F) Image 3 processed to capture ring edges.

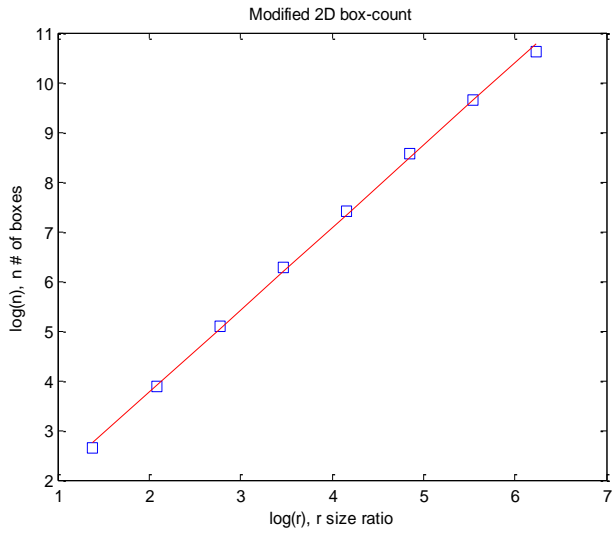

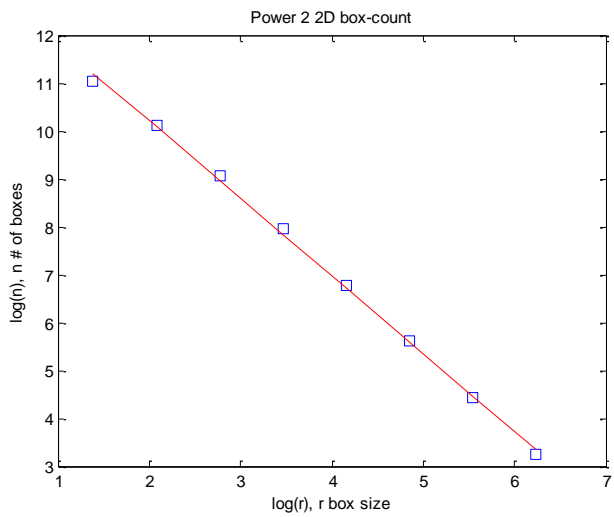
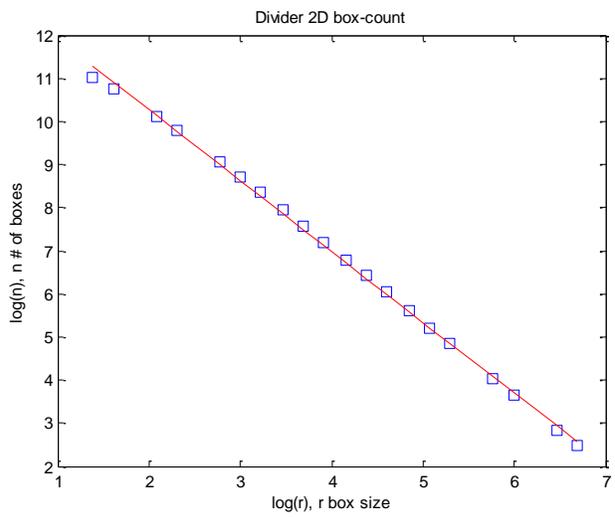

Fig. 2. Box counting method to estimate the fractal dimension of image 1: (A) Modified box counting; (B) Power 2 box counting; (C) Divider box counting.

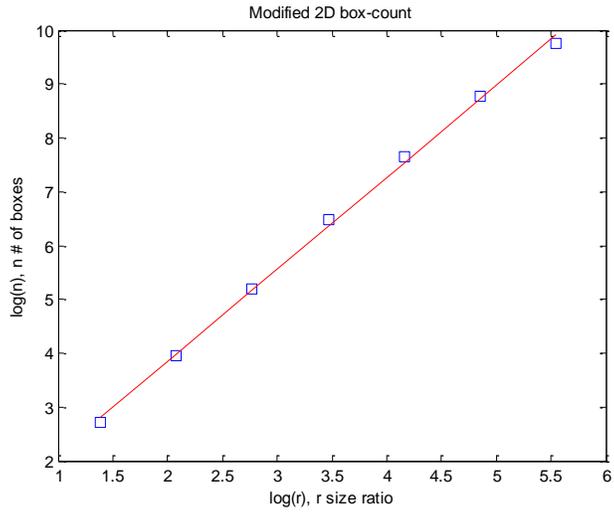

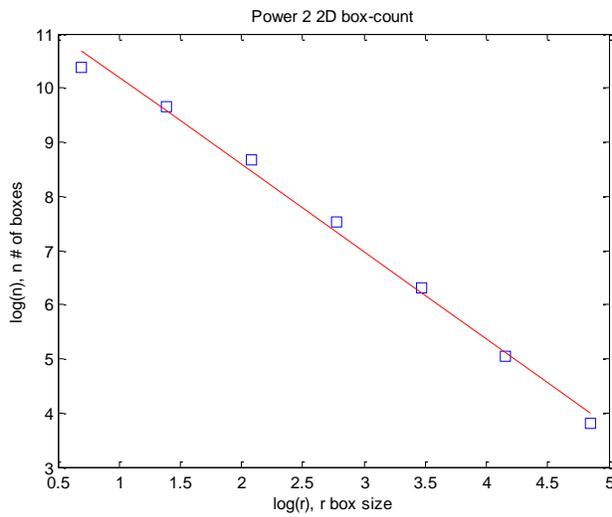
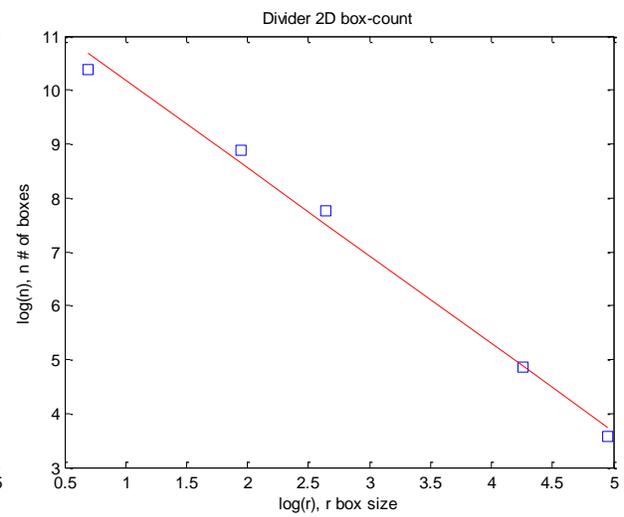

Fig. 3. Box counting method to estimate the fractal dimension of image 2: (A) Modified box counting; (B) Power 2 box counting; (C) Divider box counting.

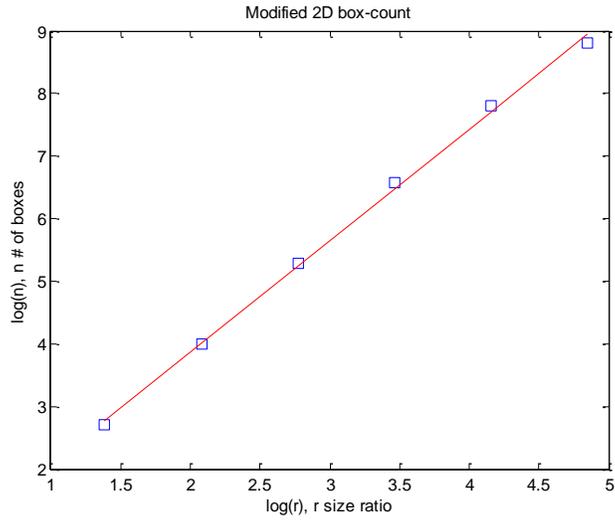

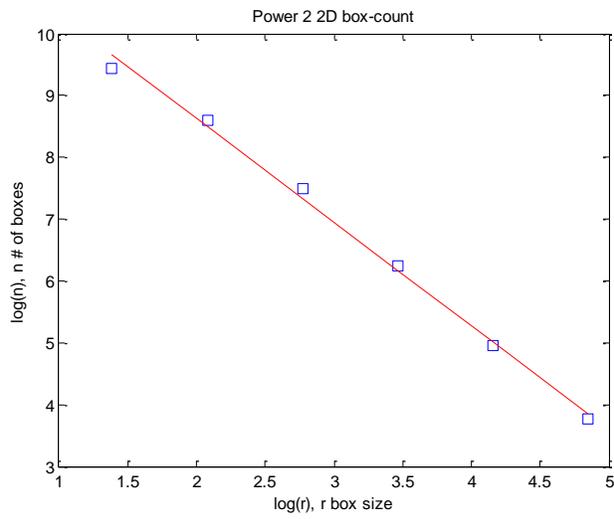
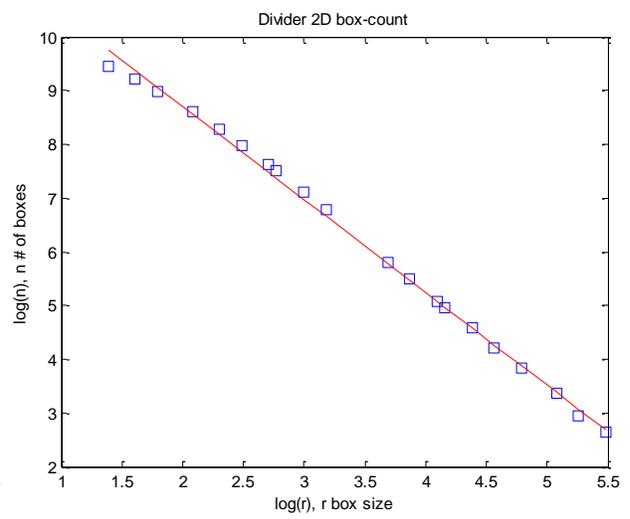

Fig. 4. Box counting method to estimate the fractal dimension of image 3: (A) Modified box counting; (B) Power 2 box counting; (C) Divider box counting.

Note that these images were projections of Saturn's rings from different angles. Following the arguments presented in (5-6), given the fact that the rings' thickness is extremely small compared to their radii, the projection onto the plane of the photograph does not affect the fractal dimension. Overall, the box counting results of all images are:

| Image Sources | Modified box counting | Power 2 box counting | Divider box counting |
|---|---|---|---|
| Fig. 1. (B) | 1.63 | 1.65 | 1.66 |
| Fig. 1. (D) | 1.64 | 1.65 | 1.71 |
| Fig. 1. (F) | 1.67 | 1.72 | 1.77 |

These images always yield fractal dimensions ~1.6 to 1.7, a convincing verification of image authenticity and our fractal dimension estimation. The fact that the rings are fractal provides one more hint to understanding the intricate mechanics and physics governing these structures of granular matter. Interestingly, somewhat related studies (7-9) found average fractal dimension ~1.7 for the projected fractal dimension of the distribution of star-forming sites (HII regions) in a sample of 19 spiral galaxies.

**Acknowledgment:** This work was made possible by the NCSA at University of Illinois and the NSF support under the grant CMMI-1030940.